\documentclass{article}
\usepackage{spconf,graphicx}
\usepackage{amsmath,amssymb}
\usepackage{url,lipsum,cite}
\usepackage{multicol,multirow}
\usepackage{xpatch,xcolor}
\usepackage{tabularx}
\usepackage{makecell, booktabs}
\newcolumntype{Y}{>{\centering\arraybackslash}X}
\usepackage{hyperref}

\hypersetup{
    colorlinks=true,
    linkcolor=purple,
    filecolor=magenta,      
    urlcolor=purple,
}

\newcommand{\newpara}[1]{\vspace{4pt}\noindent\textbf{#1}}

\title{Large-scale learning of generalised representations\\for speaker recognition}
\name{
  \begin{tabular}{c}
  Jee-weon Jung$^1$, Hee-Soo Heo$^1$, Bong-Jin Lee$^1$, Jaesong Lee$^1$,\\
  Hye-jin Shim$^2$, Youngki Kwon$^1$, Joon Son Chung$^3$, Shinji Watanabe$^4$
  \end{tabular}
}
\address{
  $^1$Naver Corporation, South Korea
  $^2$School of Computing, University of Eastern Finland, Joensuu, Finland\\
  $^3$Korea Advanced Institute of Science and Technology, South Korea\\
  $^4$Carnegie Mellon University, Pittsburgh, PA, USA
 }
\begin{document}
\ninept
\maketitle

\begin{abstract}
The objective of this work is to develop a speaker recognition model to be used in diverse scenarios. 
We hypothesise that two components should be adequately configured to build such a model. 
First, adequate architecture would be required.
We explore several recent state-of-the-art models, including ECAPA-TDNN and MFA-Conformer, as well as other baselines. 
Second, a massive amount of data would be required.
We investigate several new training data configurations combining a few existing datasets. 
The most extensive configuration includes over 87k speakers' 10.22k hours of speech. 
Four evaluation protocols are adopted to measure how the trained model performs in diverse scenarios. 
Through experiments, we find that MFA-Conformer with the least inductive bias generalises the best. 
We also show that training with proposed large data configurations gives better performance. 
A boost in generalisation is observed, where the average performance on four evaluation protocols improves by more than 20\%. 
In addition, we also demonstrate that these models' performances can improve even further when increasing capacity. 

\end{abstract}

\section{Introduction}
\label{sec:intro}
Deep neural networks (DNNs) with extremely large capacities are leading state of the art in diverse domains, including computer vision and natural language processing~\cite{dosovitskiy2020image,kenton2019bert}.
Various big models have also been studied in the audio and speech domain, including automatic speech recognition~\cite{hsu2021hubert,baevski2020wav2vec,chen2022wavlm}.
These models generalise well across various tasks and scenarios.

In general, two approaches are used to train these models. 
First, one pre-trains the model in self-supervised learning (SSL) with a massive amount of unlabelled data and then fine-tunes the model for the downstream task with collected data and corresponding labels~\cite{bao2022beit,novoselov2022robust,vaessen2022fine,fan2020exploring,chen2022large}. 
Second, a combination of multiple datasets with labels is utilised to directly train a model from scratch in a supervised manner~\cite{dosovitskiy2020image}. 
The second approach is considered costly because it requires human annotation.
However, it can also be a promising approach as the research community is witnessing more easily accessible datasets~\cite{nagrani2017voxceleb,chung2018voxceleb2,ardila2020common,chung2020spot,ryant2020third}, and hence it is the scope of this study.

In speaker recognition, there also exists several studies which leverage a pre-trained SSL model, typically wav2vec 2.0~\cite{baevski2020wav2vec}, and fine-tune it~\cite{novoselov2022robust,vaessen2022fine,fan2020exploring,chen2022large}.
Although the exploited SSL frameworks are not explicitly designed for speaker verification, using them leads to improved performance. 
However, directly training a large speaker recognition model from scratch has yet to be explored, if there exists any. 
In our preliminary experiments, which we do not have enough space to convey in this paper, we could not confirm performance improvements when we merely increased either the model size or data using a state-of-the-art ECAPA-TDNN~\cite{desplanques2020ecapa} model. 

Meanwhile, speaker recognition models\footnote{Here, speaker recognition models refer to speaker embedding extractors which can be used for open-set speaker identification, speaker verification, and speaker diarisation.} are being individually developed for different scenarios.
Several challenges pursue finding optimal models which serve the best in a particular scenario~\cite{qin2020interspeech,zeinali2020sdsv,ryant2020third,richey2018voices}. 
However, without a general backbone model for speaker recognition, the efforts are scattered, which can be inefficient. 

Our objective in this paper is to train a general speaker recognition model from scratch that can be directly adopted into various scenarios.
Although this paper has not investigated, the trained model is complementary to SSL methods and can be further fine-tuned for a specific scenario. 
We hypothesise that two aspects should be adequately addressed to achieve the goal: an adequate DNN architecture and massive data from a collection of datasets.
For the architecture, we explore the recently proposed MFA-Conformer~\cite{zhang2022mfa} and other baseline models, ECAPA-TDNN~\cite{desplanques2020ecapa}, Res2NetSE50\cite{gao2019res2net}, and RawNet3~\cite{jung2022pushing} for our study. 
We combine the widely used VoxCeleb 1\&2 datasets~\cite{nagrani2017voxceleb,chung2018voxceleb2} as well as NIST SRE datasets~\cite{przybocki2007nist,martin2009nist} and also a portion of the CommonVoice dataset~\cite{ardila2020common} and compose training data configurations.
Experiments with different model capacities are also explored in addition.

We develop a model demonstrating competitive performances in diverse scenarios through extensive experiments. 
We find that:
\begin{itemize}
    \item in terms of model architecture, while CNN-dominant ECAPA-TDNN does not, MFA-Conformer, a variant of the Transformer~\cite{vaswani2017attention}, better generalises towards a wide range of scenarios.
    \item a combination of multiple datasets could be successfully used together. 
    \item more data and parameters can even further improve the performance across diverse scenarios simultaneously.
\end{itemize}

The rest of the paper is organised as follows. 
Section \ref{sec:models} describes DNN architectures.
Various data configurations for training the model are denoted in Section \ref{sec:data}. 
Evaluation protocols for testing the model in diverse scenarios are delivered in Section \ref{sec:testset}.
Section \ref{sec:experiments} addresses the experiments and corresponding results and lastly our conclusion is presented in Section \ref{sec:conclusion}.

\section{DNN architectures}
\label{sec:models}
We employ four recently proposed models: ECAPA-TDNN~\cite{desplanques2020ecapa}, MFA-Conformer~\cite{zhang2022mfa}, Res2NetSE50~\cite{gao2019res2net}, and RawNet3~\cite{jung2022pushing}. 
Three models, except MFA-Conformer, are CNN-dominant architectures with residual connections, more specifically, Res2Net backbone block-based. 
Among the three models, we focus on ECAPA-TDNN and MFA-Conformer based empirical results (see Table~\ref{tab:main_res}).

ECAPA-TDNN applies three major modifications to the original Res2NetSE. 
First, it adopts channel and context-dependent statistics pooling to aggregate frame-level features. 
Second, it aggregates all block outputs and feeds them to the channel and context-dependent statistics pooling layer. 
Third, it uses the sum of all previous block outputs as the residual connection in each block. 

On the contrary, MFA-Conformer is a Transformer~\cite{vaswani2017attention}-based architecture which focuses more on self-attention mechanism although it includes convolution layers as well. 
After a few preceding studies~\cite{safari2020self,zhu2021serialized,mary2021s,wang2022multi} which worked on adopting Transformer for speaker verification, Zhang et al. demonstrated state-of-the-art performance with MFA-Conformer, modifying the Conformer~\cite{gulati2020conformer} model. 

MFA-Conformer has two modifications compared to the original Conformer. 
First, it concatenates all layers' outputs and feeds them to the following layer. 
We suspect this is beneficial because each layer's representation can be somewhat complementary. 
Second, it applies an attentive statistics pooling~\cite{okabe2018attentive} to the concatenated encoder outputs to aggregate the frame-level features into an utterance-level feature.

\section{Data configurations}
\label{sec:data}
We focused on two properties when building training data configurations to achieve our goal: a general-purpose speaker recognition model.
First, we wanted to enlarge the dataset by combining various datasets. 
Second, we targeted the configurations to include datasets spanning as diverse domains as possible.

Table~\ref{tab:data_cfg} describes the conventional and four training data configurations we propose. 
The conventional training data configuration, \textbf{{\em v0}}, involves VoxCeleb1\&2 development sets. 
It includes 7,205 speakers' 1.2 million utterances which account for 2.69k hours.
This training data configuration is widely used in recent literature and hence used as the baseline configuration. 

The \textbf{{\em v1}} data configuration is the main configuration of this paper. 
We first include NIST SRE 2004, 2006, and 2008 datasets~\cite{przybocki2007nist,martin2009nist} to enable the trained model to cover narrow-band telephony scenario. 
We upsample all data in NIST SRE to match other datasets' sampling rate and remove silence with voice activity detection.\footnote{\label{fn1}We use the webrtcvad from \url{https://github.com/wiseman/py-webrtcvad}.} 
Four languages, German (De), Spanish (Es), French (Fr), and Italian (It), of CommonVoice\footnote{Corpus version 9.0 is used.}~\cite{ardila2020common} are added.
Since CommonVoice is collected in a crowd-source manner, it would also introduce diverse real-world recording device channels to the trained model.
We intentionally selected languages other than English to train the model with as many languages as possible. 

\textbf{{\em v2}} to \textbf{{\em v4}} data configurations in Table~\ref{tab:data_cfg} are an extension of the \textbf{{\em v1}} configuration, which we prepared after the success of the \textbf{{\em v1}} configuration. 
The \textbf{{\em v2}} data configuration adds five more languages from CommonVoice, which are Kabyle (Kab), Luganda (Lg), Russian (Ru), Belarusian (Be), and Catalan (Ca). 
When selecting these languages among many other languages in the CommonVoice, we focused on the number of utterances per speaker and the total duration. 
The five languages mentioned above are those which satisfy both enough number of utterances and total duration.
The \textbf{{\em v3}} data configuration adds a refined English partition of the MLS dataset~\cite{pratap2020mls}, where we limit each speaker's total duration to not exceed 25 minutes. 
In this configuration, we wanted to verify (i) whether adding more English data, when we have a sufficient amount already, is beneficial and (ii) whether adding an audiobook domain is beneficial.
The \textbf{{\em v4}} data configuration combines \textbf{{\em v2}} and \textbf{{\em v3}} data configurations by adding both five languages of the CommonVoice and refined English partition of the MLS dataset. 
The intent is to train a model with an unprecedented scale of supervised data.

\begin{table}[!t]
  \centering
  \small
  \caption{
    Various exploited training data configurations (spks: speakers, utts: utterances, dur: duration in hours).
  }
  \begin{tabularx}{\columnwidth}{lYYY}
    \Xhline{1pt}
    Dataset & N spks & N utts & Dur\\ 
    \Xhline{1pt}
    \multicolumn{4}{c}{\textit{\textbf{config v0 (conventional)}}}\\
    \hline\hline
    VoxCeleb1\&2 dev & 7,205 & 1,240,651 & 2.69k\\
    \hline
    Total & 7,205 & 1,240,651 & 2.69k\\
    \Xhline{1pt}
    \multicolumn{4}{c}{\textit{\textbf{config v1 (main)}}}\\
    \hline\hline
    \textbf{\textit{config v0}} & 7,205 & 1,240,651 & 2.69k\\
    +NIST SRE 04\&06\&08 & \multirow{2}{*}{3,461} & \multirow{2}{*}{496,310} & \multirow{2}{*}{0.46k}\\
    +CommonVoice & \multirow{2}{*}{40,512} & \multirow{2}{*}{1,341,621} & \multirow{2}{*}{2.03k}\\
    \hspace{0.2cm}De\&Es\&Fr\&It\\
    \hline
    Total & 51,178 & 3,078,582 & 5.18k\\
    \Xhline{1pt}
    \multicolumn{4}{c}{\textit{\textbf{config v2 (more CommonVoice)}}}\\
    \hline\hline
    \textbf{\textit{config v1}} & 51,178 & 3,078,582 & 5.18k\\
    +CommonVoice  & \multirow{2}{*}{30,191} & \multirow{2}{*}{2,538,723} & \multirow{2}{*}{3.37k}\\
    \hspace{0.2cm}Kab\&Lg\&Ru\&Be\&Ca\\
    \hline
    Total & 81,733 & 5,617,305 & 8.55k\\
    \Xhline{1pt}
    \multicolumn{4}{c}{\textit{\textbf{config v3 (add MLS)}}}\\
    \hline\hline
    \textbf{\textit{config v1}} & 51,178 & 3,078,582 & 5.18k\\
    +MLS (En, refined) & 5,490 & 403,584 & 1.67k\\
    \hline
    Total & 56,668 & 3,482,166 & 6.85k\\
    \Xhline{1pt}
    \multicolumn{4}{c}{\textit{\textbf{config v4 (the largest)}}}\\
    \hline\hline
    \textbf{\textit{config v1}} & 51,178 & 3,078,582 & 5.18k\\
    +CommonVoice  & \multirow{2}{*}{30,191} & \multirow{2}{*}{2,538,723} & \multirow{2}{*}{3.37k}\\
    \hspace{0.2cm}Kab\&Lg\&Ru\&Be\&Ca\\
    +MLS (En, refined) & 5,490 & 403,584 & 1.67k\\
    \hline
    Total & 87,223 & 6,020,889 & 10.22k\\
    \Xhline{1pt}
  \end{tabularx}
  \vspace{-10pt}
  \label{tab:data_cfg}
\end{table}
\begin{table*}[!t]
  \centering
  \small
  \caption{
    Main results. 
    \textbf{\textit{config v0}} refers to the widely used VoxCeleb training configuration and \textbf{\textit{config v1}} describes our expanded training configuration. 
    Boldface depicts better performance comparing v0 and v1. 
    Performance reported in EER (\%, \textit{lower is better}).
  }
  \begin{tabularx}{\linewidth}{l|YY|YY|Y|Y}
    \Xhline{1pt}
    \textbf{Architecture} & \textbf{ECAPA-TDNN} & \textbf{ECAPA-TDNN} & \textbf{Conformer} & \textbf{Conformer} & \textbf{Res2NetSE50} & \textbf{RawNet3}\\ 
    \textbf{Data config} & \textbf{\textit{v0}} & \textbf{\textit{v1}} & \textbf{\textit{v0}} & \textbf{\textit{v1}} & \textbf{\textit{v0}} + MLS & \textbf{\textit{v0}}\\
    \Xhline{1pt}
    Vox1-O & \textbf{0.77} & 1.48 & 0.74 & \textbf{0.70} & 1.57 & 0.89\\
    DIHARD3 eval & 3.85 & \textbf{2.74} & 3.70 & \textbf{2.64} & 5.34 & 3.79\\
    VoxConverse test & 4.52 & \textbf{4.46} & 4.27 & \textbf{3.87} & 4.11 & 4.12\\
    NIST SRE 2010 & \textbf{11.35} & 12.09 & 9.62 & \textbf{8.70} & 16.87 & 11.29\\
    \hline
    Average & \textbf{5.12} & 5.19 & 4.58 & \textbf{3.97} & 6.97 & 5.02\\
    \Xhline{1pt}
  \end{tabularx}
  \vspace{-5pt}
  \label{tab:main_res}
\end{table*}

\section{Evaluation}
\label{sec:testset}
In order to measure how well the trained model generalises towards various scenarios, we employ four evaluation protocols. 
Two evaluation sets, DIHARD3 evaluation~\cite{ryant2020third} and VoxConverse test~\cite{chung2020spot}, are originally speaker diarisation evaluation datasets; however, we use the adapted version for speaker verification~\cite{jung2023evaluation}.\footnote{These two evaluation protocols will be made publicly accessible.}
Generating pairs using segments within each session, these two datasets mimic the speaker diarisation scenario where negatives are extremely hard because all features other than the speaker identity are the same.
We only select single-speaker segments with no overlaps and do not limit the duration. 
Using these protocols, we measure equal error rates (EERs), adopted as the metric throughout this paper.

\newpara{VoxCeleb1.} The VoxCeleb1 test set~\cite{nagrani2017voxceleb}, also known as Vox1-O, consists of 40 speakers' utterances with 37,611 trials.
The protocol is {\em in the wild} in that it involves various ethnicities, accents, channels and ages uploaded to YouTube.  
It is a widely used benchmark protocol where we use the cleaned version.

\newpara{DIHARD3 eval.} This evaluation protocol is adapted from the DIHARD3 evaluation set~\cite{ryant2020third}.
The dataset includes a wide range of domains, including meetings, restaurant conversations, and audiobooks.
The protocol consists of 426,198 trials, including 243,738 target and 182,460 non-target trials.

\newpara{VoxConverse test.} This evaluation protocol is adapted from the VoxConverse test set (v0.0.2)~\cite{chung2020spot}. 
It also includes diverse domains such as political debates, celebrity interviews, and talk shows.
It comprises 226,186 trials, including 85,452 target and 140,734 non-target trials.

\newpara{NIST SRE 2010.} We employ the NIST SRE 2010 evaluation protocol~\cite{martin2010nist} to measure how well the developed model performs in a telephone conversation scenario. 
Specifically, we use the 10sec-10sec protocol.
We upsample all data from 8kHz to 16kHz to evaluate in the same pipeline.
We also apply voice activity detection (VAD) to remove silent regions$^2$ since this dataset has substantial silent regions, whereas other datasets are speech-dominant.
The protocol comprises 54,720 trials including 540 target and 54,180 non-target trials.

\section{Experiments}
\label{sec:experiments}

\subsection{Model and training configurations}
\label{ssec:exp_cfg}
Four DNN architectures are employed in our experiments: ECAPA-TDNN, MFA-Conformer, Res2NetSE50, and RawNet3.
Unless mentioned otherwise, the below configurations are applied to all experiments when training these four DNNs. 
We use the additive angular margin softmax~\cite{deng2019arcface} with a margin of 0.3 and a scale of 30.
We apply MUSAN~\cite{snyder2015musan} noise, room impulse response~\cite{ko2017study}, playback speed, and frequency drop augmentations. 
We set the speaker embedding dimensionality to 256. 
We adopt Adam~\cite{kingma2015adam} as the optimiser with warm restarts~\cite{loshchilov2016sgdr} where the learning rate is between 1e-3 and 5e-6. 
The batch size is 200, where all utterances are from different speakers. 

\newpara{ECAPA-TDNN.} We use our implementation with 14.85 million parameters without the classification head. 
We feed 80-dimensional mel-frequency cepstral coefficients to the model.
Other hyper-parameters are identical to the 1024-channel version of the original ECAPA-TDNN~\cite{desplanques2020ecapa}.

\newpara{MFA-Conformer.} Our implementation of the MFA-Conformer models have 512-dimensional 6 layers with 8 attention head; we adopt 2,048 units for the linear layer in the feed-forward module; we set kernel size of 15 for the convolution module. 
The resulting model has 51.28 million parameters.
MUSAN noise and room impulse noise augmentation are applied during the training phase. 
AdamW~\cite{loshchilov2019decoupled} is adopted as the optimiser with a cosine annealing scheduler and 5k warm-up iterations where the learning rate varies between 1e-3 and 1e-8.
We feed 80-dimensional log mel-spectrograms to the model. 

\newpara{Res2NetSE50}. This model has the identical architecture to the original model~\cite{gao2019res2net} except the pooling layer is replaced by self-attentive pooling~\cite{cai2018exploring}. 
It has 9.29 million parameters and inputs 64-dimensional log mel-spectrograms.

\newpara{RawNet3.} also has the identical architecture with the original model~\cite{jung2022pushing} and has 16.28 million parameters. 
We adopt a stride size of 10 for the parameterised analytic filterbank layer. 
The model directly inputs raw waveforms. 

\subsection{Results and analysis}
\label{ssec:exp_res} 

\newpara{Main results.} Table~\ref{tab:main_res} addresses main results. 
The first four columns compare ECAPA-TDNN and MFA-Conformer, trained with data configurations \textbf{{\em v0}} and \textbf{{\em v1}}. 
For Res2NetSE50 and RawNet3, which are also a convolution-dominant model, we omit \textbf{{\em v1}} results because their \textbf{{\em v0}} performances are worse than ECAPA-TDNN. 

Comparing the first and the third columns, where both ECAPA-TDNN and Conformer are trained with \textbf{{\em v0}} data configuration, we observe that ECAPA-TDNN generalises less compared to MFA-Conformer when evaluated on diverse scenarios.
Although their performances on the popular Vox1-O evaluation set are similar (0.77\% and 0.74\%), which are also in line with reported performances of other papers in the literature~\cite{desplanques2020ecapa,zhang2022mfa}, when speaker diarisation and telephony scenarios are taken into account, MFA-Conformer outperforms ECAPA-TDNN by more than 20\%. 

The two models' results also differ when training with extended \textbf{{\em v1}} data configuration. 
ECAPA-TDNN seems to have relatively limited generalisation power. 
When comparing the first two columns, the performance of DIHARD3 and VoxConverse improved; however, the other two results were degraded. 
The performance of NIST SRE 2010 degrading, even though \textbf{{\em v1}} additionally includes similar domain data compared to \textbf{{\em v0}}, shows the limitation of ECAPA-TDNN's generalisability. 
On average, performance decreased by 5\%.\footnote{As we mentioned in Section~\ref{sec:intro}, ECAPA-TDNN did not benefit from an increase in the number of parameters.}

Conversely, MFA-Conformer generalised surprisingly well when fed with more data. 
Comparing columns three and four, all performances have improved. 
Improvement in Vox1-O was not expected because, comparing data configurations \textbf{{\em v0}} and \textbf{{\em v1}}, we have added other domain data, making the model less overfitted towards the VoxCeleb domain. 
We analyse that improvement on the Vox1-O, simultaneously accompanied by improvements in other scenarios, is evidence of genuine generalisation.
In addition, it is also encouraging that the performance on NIST SRE 2010 also improves in Conformer, whereas that of ECAPA-TDNN decreases, again showing the improved generalisation of the model. 
On average, with more data (\textbf{{\em v0}} to \textbf{{\em v1}}), MFA-Conformer's performance further improved by 13\%. 
In our thoughts, these results could have occurred because MFA-Conformer poses less inductive bias, relying on self-attention mechanisms than convolution layers.
It may mean that attention-based architectures generalise better than convolution-based architecture when abundant data is available.
Hereafter, we focus on MFA-Conformer and experiment with other aspects. 

\begin{table}[!t]
  \centering
  \small
  \caption{
    MFA-Conformer with different training data configurations. 
    Boldface depicts improved performance compared to \textbf{{\em v1}} training data configuration. 
  }
  \begin{tabularx}{\columnwidth}{l|YYYY}
    \Xhline{1pt}
    \textbf{Data config} & \textbf{\textit{v1}}& \textbf{\textit{v2}} & \textbf{\textit{v3}} & \textbf{\textit{v4}}\\ 
    \Xhline{1pt}
    Vox1-O & 0.70 & 0.80 & 0.71 & 0.78\\
    DIHARD3 eval & 2.64& \textbf{2.57} & \textbf{2.46} & \textbf{2.42}\\
    VoxConverse test & 3.87& \textbf{3.77} & 3.87 & 3.97\\
    NIST SRE 2010 & 8.70&8.94 & 8.70 & \textbf{8.36}\\
    \hline
    Average & 3.97& 4.02 & \textbf{3.93} & \textbf{3.88}\\
    \Xhline{1pt}
  \end{tabularx}
  \vspace{-10pt}
  \label{tab:var_cfg_res}
\end{table}
\begin{table}[!t]
  \centering
  \small
  \caption{
    Configurations for ablation studies regarding the model size of MFA-Conformer.
    Results are addressed in Table~\ref{tab:ablation}.
  }
  \begin{tabularx}{\columnwidth}{lcY}
    \Xhline{1pt}
    \textbf{Data config} & \textbf{Name} & \textbf{Modifications}\\ 
    \Xhline{1pt}
    v1 & Deeper & layer 6$\rightarrow$12\\
    \hline
    \multirow{2}{*}{v1} & \multirow{2}{*}{Wider} & dimensionality 512$\rightarrow$1,024\\
    & & attention head 8$\rightarrow$16\\
    \hline
    v2 & Batch size & 200$\rightarrow$400\\
    \hline
    \multirow{2}{*}{v2-v4} & \multirow{2}{*}{Wider} & dimensionality 512$\rightarrow$768\\
    & & attention head 8$\rightarrow$12\\
    \hline
    v2-v4 & Embedding & speaker embedding 256$\rightarrow$512\\
    \Xhline{1pt}
  \end{tabularx}
  \vspace{-10pt}
  \label{tab:abl_cfg}
\end{table}

\newpara{Various training configurations.}
Table~\ref{tab:var_cfg_res} and Table~\ref{tab:ablation} deliver the results with additional training data configurations: \textbf{{\em v2}}, \textbf{{\em v3}}, and \textbf{{\em v4}} from Table~\ref{tab:data_cfg} using MFA-Conformer. 
Table~\ref{tab:abl_cfg} describes the configurations for results in Table~\ref{tab:ablation}.
Through Table~\ref{tab:var_cfg_res}, we find that adding further data to configuration \textbf{{\em v1}} can bring further improvement on average without modifying the number of parameters. 
Compared to MFA-Conformer's performance on \textbf{{\em v1}} configurations (3.97\%), both \textbf{{\em v3}} and \textbf{{\em v4}} configurations brought further improvements.
However, improvements across each dataset were inconsistent, where performance on only one or two datasets improved.
Also, merely adding more CommonVoice (\textbf{{\em v2}}) did not improve the performance.

We further explore ablation experiments by increasing the models' capacity in several ways and report the result in Table~\ref{tab:ablation}.
An underlying assumption here is that because the training data increased, the model's capacity (i.e., the number of parameters) should also increase.
We did not observe meaningful improvements through experiments on data configuration \textbf{{\em v1}} to \textbf{{\em v3}}.
For configuration \textbf{{\em v1}}, we experimented with going wider and deeper by increasing the dimensionality or having more layers, where both results did not show improvements.
For configuration \textbf{{\em v2}}, we experimented with three ablations; however, only minor improvements were observed (4.02\% to 3.99\% for Embedding), and improvements were inconsistent across different domains. 
Configuration \textbf{{\em v3}} also did not give good results. 

Training data configuration \textbf{{\em v4}}, the largest one with 88k speakers, demonstrated better results.
Increasing the dimensionality of speaker embedding showed improvement in two evaluation protocols where only NIST SRE 2010's performance decreased by 0.15\% absolute. 
Further improvements were observed when we increased each layer's dimensionality and speaker embedding's dimensionality simultaneously (Wider\&Embedding), demonstrating the best performance throughout this paper. 
This result is meaningful because compared to plain \textbf{{\em v4}} configuration in Table~\ref{tab:var_cfg_res}, all performances except DIHARD3 have improved, and the performance on DIHARD3 was only degraded by 0.03\% absolute. 
Also, results on the Vox1-O evaluation protocol improved, which signifies actual generalisation because the portion of VoxCeleb data in the training configuration decreases.
Comparing this result to the \textbf{{\em v1}} configuration, we observe improvements in all four evaluation protocols. 
Compared to the \textbf{{\em v0}} configuration, the average improvement is over 20\%.
We conclude that training with more data and a bigger model can further improve generalisation across diverse domains.

\newpara{Additional remarks.}
We also experimented using the original Transformer \cite{vaswani2017attention}.
However, EER was over 2\% on Vox1-O, similar to preceding works~\cite{safari2020self,zhu2021serialized,mary2021s,wang2022multi}.
Thus, we selected MFA-Conformer instead to explore a self-attention-based architecture.
 
\begin{table}[!t]
  \centering
  \small
  \caption{
    Ablation experiments on the depth, width, and speaker embedding dimensionality using MFA-Conformer. Improved performances compared to the corresponding models in Table~\ref{tab:var_cfg_res} are depicted in boldface.
  }
  \begin{tabularx}{\columnwidth}{l|YY|Y}
    \Xhline{1pt}
    & \multicolumn{2}{c|}{\multirow{2}{*}{\textit{\textbf{config v1 (main)}}}} & \textbf{\textit{config v3}}\\
    & & & \textbf{\textit{(add MLS})}\\
    \hline\hline
    & \textbf{Deeper} & \textbf{Wider} & \textbf{Embedding}\\
    \hline
    Vox1-O & 0.74 & 0.80 & 0.85\\
    DIHARD3 eval & 2.77 & 2.72 & 2.53\\
    VoxConverse test & 4.01 & 4.09 & \textbf{3.86}\\
    NIST SRE 2010 & \textbf{8.51} & 9.44 & 9.15\\
    \hline
    Average & 4.00 & 4.26 & 4.10\\
    \Xhline{1pt}
  \end{tabularx}
  \begin{tabularx}{\columnwidth}{l|YYY}
    \multicolumn{4}{c}{\textit{\textbf{config v2 (more CommonVoice)}}}\\
    \hline\hline
    & \textbf{Wider} & \textbf{Embedding} & \textbf{Batch size}\\
    \hline
    Vox1-O & \textbf{0.72} & \textbf{0.74} & \textbf{0.74}\\
    DIHARD3 eval & 2.61 & 2.69 & 2.64\\
    VoxConverse test & 3.97 & 3.80 & 3.79\\
    NIST SRE 2010 & \textbf{8.70} & \textbf{8.74} & 9.11\\
    \hline
    Average& \textbf{4.00} & \textbf{3.99} & 4.07\\
    \Xhline{1pt}
    \multicolumn{4}{c}{\textit{\textbf{config v4 (the largest)}}}\\
    \hline\hline
    & \multirow{2}{*}{\textbf{Wider}} & \multirow{2}{*}{\textbf{Embedding}} & \textbf{Wider\&}\\
    & & & \textbf{Embedding}\\
    \hline
    Vox1-O & \textbf{0.76} & \textbf{0.69} & \textbf{0.68}\\
    DIHARD3 eval & 2.48 & 2.42 & 2.45\\
    VoxConverse test & \textbf{3.93} & \textbf{3.81} & \textbf{3.85}\\
    NIST SRE 2010 & 9.16 & 8.51 & \textbf{8.33}\\
    \hline
    Average & 4.08 & \textbf{3.85} & \textbf{3.82}\\
    \Xhline{1pt}
  \end{tabularx}
  \vspace{-10pt}
  \label{tab:ablation}
\end{table}

\section{Conclusion}
\label{sec:conclusion}
We developed a general-purpose speaker recognition model where a single model can be utilised for diverse scenarios. 
The developed model can be used in speaker verification and diarisation, where it can also process upsampled 8kHz telephony inputs.
Diverse DNN architectures, both convolution-dominant and self-attention-based, have been explored as well as increased training data and different model sizes.
Comparing ECAPA-TDNN and MFA-Conformer, we showed that self-attention-based MFA-Conformer generalises better when fed with more training data which spans diverse domains.
We also demonstrated that further improvements could be observed when simultaneously increasing the data and the number of parameters.
The single developed model successfully demonstrated improved performance in diverse scenarios.

\clearpage
\bibliographystyle{IEEEbib}
\bibliography{shortstrings,refs}
\end{document}